\documentclass[prd,aps,superscriptaddress,twocolumn,nofootinbib]{revtex4-2}

\usepackage{graphicx} 
\usepackage{amsmath} 
\usepackage{amssymb} 
\usepackage[caption=false]{subfig} 
\usepackage{hyperref} 
\usepackage{xcolor} 
\usepackage{physics} 
\usepackage[capitalize]{cleveref} 
\usepackage{bm}
\usepackage{tikz}

\newcommand{\cP}{\mathcal{P}}%
\newcommand{\deltac}{\delta_\mathrm{c}}
\DeclareMathOperator{\erfc}{erfc}
\DeclareMathOperator{\e}{e}

\graphicspath{{figures/}}

\begin{document}

\title{Small-scale clustering of Primordial Black Holes: cloud-in-cloud and exclusion effects}

\author{Pierre Auclair}
\email{pierre.auclair@uclouvain.be}

\affiliation{Cosmology, Universe and Relativity at Louvain (CURL), Institute of Mathematics and Physics, University of Louvain, 2 Chemin du Cyclotron, 1348 Louvain-la-Neuve, Belgium}
\author{Baptiste Blachier}
\email{baptiste.blachier@uclouvain.be}
\affiliation{Cosmology, Universe and Relativity at Louvain (CURL), Institute of Mathematics and Physics, University of Louvain, 2 Chemin du Cyclotron, 1348 Louvain-la-Neuve, Belgium}

\begin{abstract}
    Using an excursion-set approach, we revisit the initial spatial clustering of Primordial Black Holes (PBHs) originating from the Hubble reentry of large Gaussian density fluctuations in the early Universe.
    We derive the two-point correlation functions of PBHs, properly accounting for the ``cloud-in-cloud'' mechanism.
    Our expressions naturally and intrinsically correlate the formation of pairs of PBHs, which is a key difference with the Poisson model of clustering.
    Our approach effectively includes short-range exclusion effects and clarifies the clustering behaviors at small scale: PBHs are anticorrelated at short distances.
    Using a scale independent collapse threshold, we derive explicit expressions for the excess probability to find pairs of PBHs separated by a distance $r$, as well as the excess probability to find pairs with asymmetric mass ratio.
    Our framework is model independent by construction and we discuss possible other applications.
\end{abstract}

\maketitle

\section{Introduction}

Soon after Primordial Black Holes (PBHs) were proposed~\cite{Hawking:1971ei,Carr:1974nx,Carr:1975qj}, it was realized these objects could prove relevant in numerous aspects of cosmology and the early Universe.
They might constitute some or all of the dark matter content of the Universe~\cite{Chapline:1975ojl, Carr:2020gox} and seed the formation of supermassive black holes in galactic nuclei~\cite{Meszaros:1975ef, Bean:2002kx}.
More recently, the physics of PBHs has gained a renewed interest, as some binary mergers detected by LIGO/VIRGO/KAGRA might arise from the mergers of PBHs~\cite{Bird:2016dcv, Sasaki:2016jop, Clesse:2016vqa, Ali-Haimoud:2017rtz, Hutsi:2020sol, DeLuca:2021wjr, Franciolini:2021tla, Franciolini:2022tfm}.

The mass spectrum of PBHs ought to satisfy various observational constraints ranging from microlensing~\cite{Macho:2000nvd, EROS-2:2006ryy, Carr:2020gox}, cosmic microwave background (CMB) \cite{Ali-Haimoud:2016mbv, Ricotti:2007au, Poulin:2017bwe} and limits to their merger rates when confronted with the gravitational waves detector data~\cite{Clesse:2017bsw, Ballesteros:2018swv}.
However, these constraints depend heavily on their spatial clustering at formation.
Sizeable clustering could change the past and present merger rate of PBH binaries~\cite{Clesse:2017bsw, Clesse:2016vqa, Ballesteros:2018swv},
the way they generate certain cosmological structures \cite{Carr:2018rid, Raidal:2017mfl} and even relax the aforementioned CMB and microlensing bounds \cite{Garcia-Bellido:2017xvr, Calcino:2018mwh, Carr:2019kxo}; although this very last statement was recently challenged in~\cite{Petac:2022rio, Gorton:2022fyb}.
Since the evolution of PBH clustering necessarily enters a complex nonlinear regime, the amount of initial clustering can produce drastic effects on the subsequent evolution.

Consequently, a careful attention should be put on the initial small-scale clustering of PBHs.
Most approaches used in the literature rely on large-scale structure formalism~\cite{Peebles:1980} that was ubiquitously applied in the context of galaxy or halo formation.
In there, spatial clustering is described by the two-point correlation function.
Consecutive to sky surveys-based analysis (see e.g.\ \cite{Klypin1983, Bahcall1983}), early theoretical works relied on the so-called Press-Schechter (PS) formalism \cite{Press:1973iz}, which originally aimed to model the mass fraction of gravitationally formed objects.
Beyond the initial PS application, various expressions of the spatial correlation functions were obtained, with diverse range of validity~\cite{Kaiser1984, Politzer1984, Jensen1986}.
More refined methods, notably using excursion-set theory which aims at curing the so-called ``cloud-in-cloud'' drawback of the PS model (see \cref{sec:section2} for detailed explanations), were later developed and applied to study the spatial clustering of galaxies~\cite{Porciani:1998ye, Zentner:2006vw} in parallel with bias theory which draws a statistical relation between the spatial distribution of certain objects and the density fields sourcing their formation (see Ref.~\cite{Desjacques:2016bnm}).
A further refinement is to go beyond the approximation of pointlike objects and to account for their spatial extension, thus leading to exclusion effects \cite{Mo:1995cs, Baldauf:2013hka, Baldauf:2015fbu}.

In the specific context of PBHs formed out of the gravitational collapse of large density fluctuations in the early Universe, the issue of their spatial clustering was first addressed in Ref.~\cite{Chisholm:2005vm}, and then later corrected in Refs.~\cite{Ali-Haimoud:2018dau,Atal:2020igj} relying on a Press-Schechter-inspired model.
Using the peak approach to large-scale structures~\cite{Kaiser1984, Bardeen:1985tr}, spatial clustering was further explored in~\cite{Desjacques:2018wuu} where it was claimed that for narrow power spectrum, i.e. localized in terms of wave numbers, clustering is irrelevant. This statement was extended to broad spectrum in~\cite{MoradinezhadDizgah:2019wjf}, using an excursion-set approach but in a two-barrier setup, thus focusing on the formation history of PBHs, and the probability that PBHs are partitioned over time.

The purpose of the present paper is to derive explicit expressions characterizing the two-point statistics of the spatial distribution of PBHs (when the underlying density field is Gaussian) in the framework of the excursion-set theory with uncorrelated steps thus accounting for ``cloud-in-cloud'' and exclusion effects at short separation scales. Contrary to the two-barrier excursion-set problem, we only focus on the \textit{initial} clustering of PBHs, without investigating the possible time evolution of virialized structures~\cite{MoradinezhadDizgah:2019wjf}, nor the nonlinear evolution of the clusters \cite{DeLuca:2020jug}.

It is organized as follows: in \cref{sec:section2}, we present the main lines of the excursion-set approach, with a specific emphasis on how it enables to cure the so-called ``cloud-in-cloud'' issue arising from the most frequently used Press-Schechter formalism.
We also introduce joint probabilities, which are key parameters to study two-point correlations, with the aim to remain as generic as possible.
In \cref{sec:section3}, we apply these tools in the case of a scale independent threshold for PBH formation mechanism.
The result of \cref{sec:section3} can be seen as the most natural excursion-set extension of the two-point correlation function derived in Ref.~\cite{Ali-Haimoud:2018dau}, and we show in \cref{sec:small} that it readily includes short-range exclusion effects, leading to a better and well-defined comprehension of clustering behaviors at small scale.
\cref{sec:section4} derives novel analytical expressions for correlations at the level of pairs of PBHs.
As our approach remains agnostic of the details of the PBH formation -- in particular independent of the shape of the density field power spectrum -- we discuss in \cref{sec:section5} the scope of the possible use of the various spatial correlations introduced herein.

\section{Excursion-set formalism}
\label{sec:section2}

\subsection{Motivations}

The prototypical scenario of PBH formation assumes they originate from large density fluctuations which reenter the Hubble radius and collapse into black holes~\cite{Sasaki:2018dmp}.
The cosmological fluctuations that are expected to lead to PBH collapses are described by an overdensity field $\delta (\vb{x})$, with known statistical properties, e.g.\ a Gaussian random field.
A region in which the density contrast is larger than some threshold value $\deltac$ forms a PBH. Let us note that the statistics of PBHs are exponentially sensitive to this threshold of formation \cite{Carr:1975qj}. Its physical nature was studied in Refs. \cite{Shibata:1999zs, Musco:2018rwt}. Giving criteria to determine it with precision is a difficult task: numerical simulations were proposed \cite{Jedamzik:1999am, Shibata:1999zs,Hawke2002,Nakama:2013ica, Escriva:2019nsa} as well as analytical estimates \cite{Carr:1975qj, Harada:2013epa, Escriva:2019phb, Escriva:2020tak, Musco:2020jjb} (see Ref. \cite{Escriva:2021aeh} for a generic review).

\begin{figure}
\begin{center}
    \begin{tikzpicture}
        \draw [very thick] [->] (0,3) -- (6,3);
        \draw  (6,3) node [right] {$R$};
        \draw (3,3) node [above left] {$R$};
        \draw [very thick] [->] (0,2) -- (6,2);
        \draw  (6,2) node [right] {$M(R)$};
        \draw [very thick] [<-] (0,1) -- (6,1);
        \draw (6,1) node [right] {$S(R)$};
        \draw (3,1) node [above right] {$S$};
        \draw (3,0) node [below] {$\delta_{R}$};
        \draw [dotted] (3,0) -- (3,4);
    \end{tikzpicture}
\end{center}
    \caption{There is a one-to-one relation between the variance $S$, the size of the coarse graining $R$ and the mass of the produced PBH $M$, which can be used interchangeably, provided a choice is made for the power spectrum. The arrow shows increasing values of the given variable.}
    \label{fig:scales}
\end{figure}
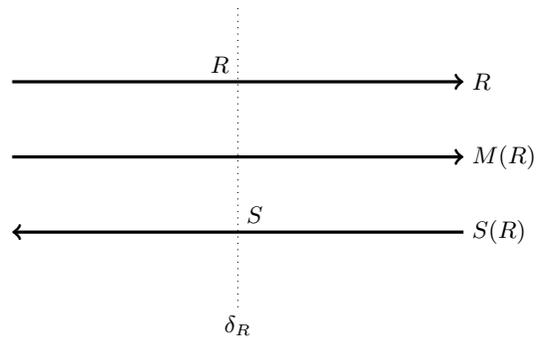

In practice, we define the coarse-grained density perturbation over a spherical region of radius $R$ about a point $\vb{x}$
\begin{equation}
    \delta_{R}(\vb{x}) = \int \frac{\dd[3]{\vb{k}}}{(2\pi)^{3}} \delta(\vb{k}) W(k,R) \e^{-i\vb{k} \cdot \vb{x}}
\end{equation}
where $W$ is a window function selecting a subset of wave numbers.
For practical reasons, we usually take $W$ to be a top-hat function in Fourier space.
Assuming Gaussian statistics for $\delta$, so does $\delta_{R}$ and the probability density associated to $\delta_{R}$ reads as
\begin{equation}
    P(\delta_{R}) = \frac{1}{\sqrt{2\pi S(R)}} \exp[-\dfrac{\delta_{R}^{2}}{2 S(R)}],
\end{equation}
which is entirely determined by its variance $S(R) \equiv \ev{\delta_{R}^{2}(\vb{x})}$.
An important remark is that if one introduces the (reduced) power spectrum of $\delta$, $\cP_{\delta}$, the variance $S$ can be computed through
\begin{equation}
    S(R) = \int_{0}^{\infty} \cP_{\delta}(k) W^{2}(k,R) \dd{\ln k}.
\label{eq:S_and_R}
\end{equation}
The above equation relates the variance $S$ to the smoothing scale $R$ in a monotonous way: $S(R)$ is decreasing with $R$.
Furthermore, there is a one-to-one correspondence between a structure of scale $R$ and its associated mass $M$ since, at leading order in perturbations, $M = 4 \pi \overline{\rho} R^{3}/3$, with $\overline{\rho}$ the energy density of the background.
Hence, in principle, one could use $M$, $R$ and $S$ interchangeably to describe the scale of PBH formation, as illustrated in \cref{fig:scales}.
In the rest of the paper, we mainly work with the variance $S$, as it can be interpreted as a ``time'' variable and enables us to remain agnostic about the model of PBH formation.
To apply our results to physical models, the correspondence between $S$, $R$, and $M$ should be established using \cref{eq:S_and_R}.

A first approach to determine the probability of forming a gravitationally bound structure is the Press-Schechter (PS) formalism \cite{Press:1973iz}.
Under this framework, it is assumed that a region of scale $R$ has collapsed on a scale $R' > R$ if and only if the coarse-grained density contrast is above the threshold.
In terms of probability, it reads
\begin{equation}
    P\left[ \delta_{R}>\deltac(R) \right]
    = \int_{\delta_c}^\infty P(\delta_r) \dd{\delta_r}
    = \frac{1}{2} \erfc \left[ \frac{\deltac(R)}{\sqrt{2 S(R)}} \right]
\end{equation}
where $\erfc$ is the complementary error function.
It is well-known that the PS approach is plagued by a miscount of collapsed objects because of the so-called ``cloud-in-cloud'' problem.
Indeed, the Press-Schechter formalism assumes that a region of radius $R$ collapses as soon as $\delta_{R}$ exceeds a certain threshold $\deltac(R)$.
This procedure misses the cases in which, on a given smoothing scale $R$, $\delta_{R}$ is below the threshold, but it still happened to be above the threshold at some scale $R' > R$ (which corresponds to structures of mass $M' > M$).

The analysis of the spatial clustering performed in Ref.~\cite{Ali-Haimoud:2018dau} relies on such a formalism and therefore also suffers from the ``cloud-in-cloud'' problem, as we detail in \cref{sec:section3}.

\subsection{``Cloud-in-cloud'' resolution}

The excursion-set method aims to cure this ``cloud-in-cloud'' -- see Refs.~\cite{Peacock:1990zz, Bower:1991kf, Bond:1990iw} for pioneering works on the subject and Refs.~\cite{redner_2001, Maggiore:2009rv, Auclair:2020csm} for more recent reviews.
In this approach the density perturbation $\delta_R(\vb{x})$ evolves stochastically with the smoothing scale $R$, or equivalently with $S$.
Going from one scale $R$ to a smaller scale $R - \Delta R$, the density perturbation is modified by the addition of new modes $\delta(\vb{k})$.
The probability to form a PBH about $\vb{x}$ and its scale are mapped into a first-passage-time problem in the presence of a barrier $\deltac(S)$.
The ``cloud-in-cloud'' issue is avoided by counting only the largest gravitationally bound structures, i.e.\ a PBH.

In practice, the coarse-grained density field $\delta_{R}$ is promoted to a stochastic quantity whose evolution in terms of its variance $S$ (playing the role of the ``time'' variable) obeys a Langevin equation with a white Gaussian noise $\xi(S)$
\begin{equation}
    \dv{\delta_R}{S} = \xi(S).
\end{equation}
The evolution of $\delta_{R}$ can be seen as a random trajectory submitted to a Brownian process.
The fact that the noise is white arises from the specific choice of a top-hat window function in Fourier space to coarse-grain the density field.
This enforces the Markovian (i.e.\ fully uncorrelated) nature of the steps of the random walk.\footnote{Non-Markovian excursion-set approaches result in coloured noise in the Langevin equation, making the analysis more involved, see e.g.\ \cite{Musso:2013pha, Nikakhtar:2018qqg}.}
In this context, the probability distribution $P$ of $\delta_{R}$ can be determined by solving the adjoint Fokker-Planck equation
\begin{equation}
    \pdv{P}{S} = \frac{1}{2} \pdv[2]{P}{\delta_{R}}\ , \quad P(\delta_R, S_i) = \delta_D(\delta_R - \delta_i) \ ,
\label{eq:Fokker-Planck}
\end{equation}
with $\delta_D$ the Dirac distribution and $(\delta_i, S_i)$ serving as initial conditions for the trajectory.

The solution, in absence of any boundary condition which we denote with a subscript ``free'', is a Gaussian:
\begin{equation}
    \label{eq:free}
    P_{\mathrm{free}} (\delta_{R}, S \vert \delta_i, S_i)
    = \frac{1}{\sqrt{2\pi(S-S_i)}} \exp \left[- \frac{(\delta_{R} - \delta_i)^{2}}{2(S-S_i)}\right].
\end{equation}
It represents the probability density that the coarse-grained density contrast takes the value $\delta_{R}$ at time $S$, given that at ``initial'' time $S_i$, its value is $\delta_i$.
For notation convenience, in the rest of this paper, we shall drop the index $R$ in $\delta_R$, but one should keep in mind that we work with a coarse-grained density field.
Formally, this $P_{\mathrm{free}}$ is the assumption made in the Press-Schechter approach, assuming a homogeneous field on very large scales, i.e.\ $(\delta_i, S_i) = (0,0)$.
In this picture, the ``cloud-in-cloud'' problem arises because this formalism allows multiple crossings for a trajectory.

To remediate this issue, one should only consider first up-crossings with the threshold $\deltac(S)$, which is equivalent to calculating the lowest value of $S$ for which the trajectory crosses the threshold.
Formally, it boils down to considering the solution $P(\delta, S \vert \delta_i, S_i)$ of \cref{eq:Fokker-Planck} with an absorbing boundary at $\delta = \deltac(S)$.
It represents realizations of the Langevin equation beginning at $(\delta_i, S_i)$ that, at time $S$, have not yet crossed the absorbing boundary.
The scale of the formed PBH is given by $P_{\mathrm{FPT}}(s \vert \delta_i, S_i)$, the probability density of the first-passage-time of the boundary $\deltac(S)$ starting from $(\delta_i, S_i)$\footnote{We consider here in full generality \textit{conditional} probabilities, we do not fix the initial condition to $\delta = 0$ for $S=0$.
Nonetheless, identical computations to the one performed in Ref. \cite{Auclair:2020csm} could be carried out.}.

From the condition that, at a given ``time'' $S$, any realization of the Langevin equation has either crossed out the threshold at a previous time $s<S$, or still contributes to the distribution $P$
\begin{equation}
    1 = \int_{S_i}^{S} \dd{s} P_{\mathrm{FPT}}(s \vert \delta_i, S_i) + \int_{-\infty}^{\deltac(S)} P(\delta, S \vert \delta_i, S_i) \dd{\delta},
\end{equation}
one can obtain by differentiating with respect to $S$ the following implicit expression for the (conditional) first-passage-time distribution
\begin{widetext}
\begin{multline}
     P_{\mathrm{FPT}}(S \vert \delta_i, S_i) = \left[\frac{\deltac(S) - \delta_i}{S-S_i} - 2 \deltac'(S) \right] P_{\mathrm{free}}\left[\deltac(S), S\vert \delta_i, S_i\right] \\
     + \int_{S_i}^{S} \dd{s} \left[2 \deltac'(S) - \frac{\deltac(S) - \deltac(s)}{S-s} \right] P_{\mathrm{free}} \left[\deltac(S), S\vert\deltac(s), s\right] P_{\mathrm{FPT}}(s \vert \delta_i, S_i).
\label{eq:P_FPT_generic}
\end{multline}
\end{widetext}
In principle, this integro-differential equation for arbitrary $\deltac(S)$ can be solved numerically. The singularity appearing in the integrand of the second term when $s \to S$ can lead to numerical instabilities. Ref.~\cite{Auclair:2020csm} shows that the introduction of a kernel $K(s)$, set to take the value $\deltac'(s)$, can greatly improve precision.

If the formation threshold is scale independent, the above equation greatly simplifies and this is the case we will consider in \cref{sec:section3,sec:section4}.
Let us also remark that we recover exactly the results of \cite{Auclair:2020csm} when taking the limit $S_{i} \to 0, \delta_{i} \to 0$.
Otherwise stated, considering conditional probability boils down to performing a constant shift from $(\delta = 0, S= 0)$ to the desired initial values $\delta_{i}$ of the field at the initial chosen ``time'' $S_{i}$.

Eventually, we define a shorter notation for the first-passage-time distribution in the limit $S_{i} \to 0, \delta_{i} \to 0$:
\begin{equation}
    P_{\mathrm{FPT}}(S) \equiv P_{\mathrm{FPT}}(S \vert 0, 0).
\end{equation}

\begin{figure*}
    \centering
    \includegraphics[width=.7\textwidth]{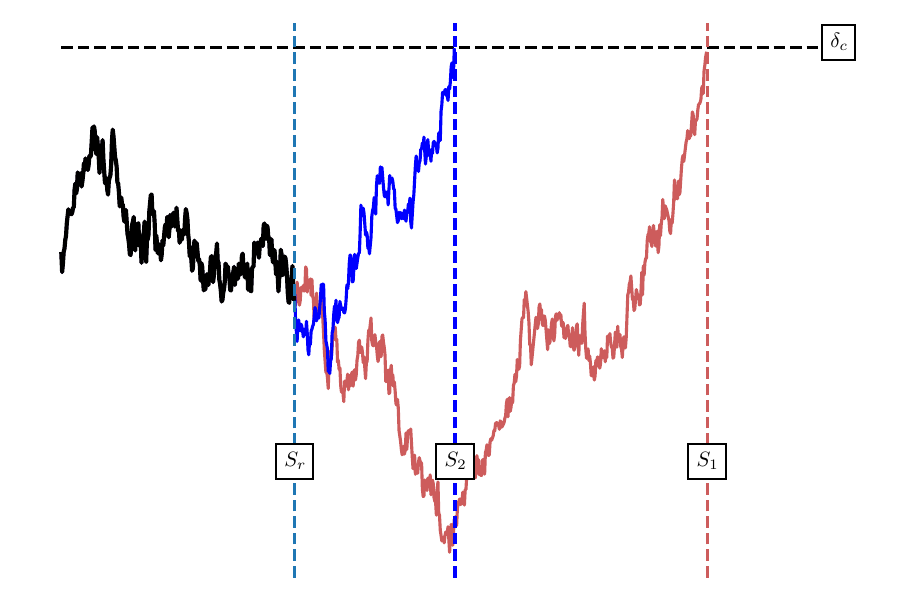}
    \caption{Schematic representation of random walks performed by smoothed overdensities sharing a common past until ``time'' $S_{r}$ and which subsequently evolve in distinct ways, resulting in two different collapses that occur respectively at the first-crossing times $S_{1}$ and $S_{2}$.}
    \label{fig:random_bundle}
\end{figure*}

\subsection{Joint probability}
\label{sec:joint}

In this section, we determine $\cP(S_{1}, S_{2};r)$, the joint probability of finding two PBHs of masses $S_1, S_2$ within a distance $r$ from each other.
Note that, up to this point, the tools we introduced are also standard in Dark Matter halos formation~\cite{Desjacques:2016bnm}. Here, we make use of an idea successfully applied to understand the clustering of stars at formation~\cite{Hopkins:2011uk, Hopkins:2012bi, Hopkins:2012fm}, and adapt it to the specificities of PBH formation.

As explained earlier, the variance $S$ is related in a monotonous way, yet model dependent, to PBH mass $M$.
Consequently, $\cP(S_{1}, S_{2};r)$ represents the probability of finding a PBH that has collapsed at a scale $R(S_1)$ with mass $M(S_1)$ at a distance $r$ from a PBH which collapsed at a scale $R(S_2)$ with mass $M(S_2)$.

\begin{figure*}
    \centering
    \includegraphics[width=.7\textwidth]{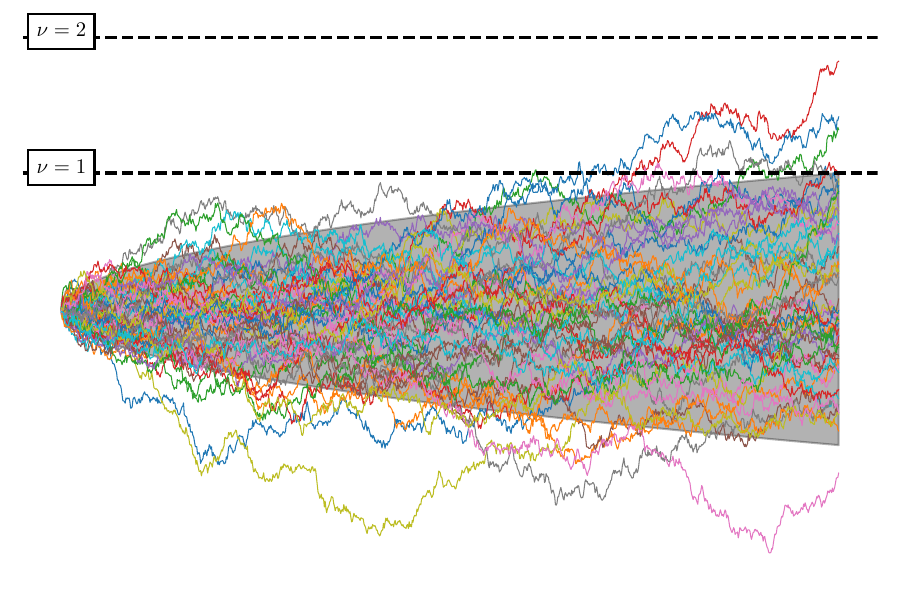}
    \caption{Schematic representation of random walks, from $S=0$ to the maximum value $\sigma^2$.
    Most of the trajectories are confined in the gray region which ranges between $\pm \sigma$.
    The horizontal lines correspond to different values of $\nu = \delta_c / \sigma$, where $\nu$ is a measure of the barrier's height in units of the typical spread of the trajectories.
    Low values of $\nu$ indicate that barrier crossings are frequent: hence multiple crossings are also frequent. Higher values of $\nu$ indicate that barrier crossings (and therefore PBH collapses) are rare events, and multiple crossings are suppressed.}
    \label{fig:random_traj}
\end{figure*}

In the excursion-set approach, to each of these black holes corresponds a trajectory $\delta(S)$ which first crosses the threshold $\deltac(S)$ at $S_n$, $n \in \{1, 2\}$.
These two collapsed structures are separated by a distance $r$, meaning their trajectories share a ``common past'' up to a certain time $S_r$, and then evolve freely from $S_r$ until their collapse at $S_n$, as illustrated in \cref{fig:random_traj}.
The main difference between Refs.~\cite{Hopkins:2011uk, Hopkins:2012bi, Hopkins:2012fm} and our work is that stars can be included in larger gravitationally-bound structures such as molecular clouds. Therefore, stars are associated with last-passage-time, instead of first-passage-time in the context of PBH formation.

We introduce a new object called \emph{backward probability} $P_{\mathrm{bw}}\left[\delta_{r}, S_{r} \vert \deltac(S_{1})\right]$, which represents the probability of a trajectory having the value $\delta_{r}$ at $S_{r}$ \textit{given} that it collapses at the scale $S_{1}$ [i.e.\ that it is such that $\delta(S_{1}) = \deltac(S_{1})$: hence the notation].
One can define the backward probability \emph{via} an argument similar to Bayes' theorem
\begin{equation}
    P_{\mathrm{bw}}\left[\delta_{r}, S_{r} \vert \deltac(S_{1})\right] P_{\mathrm{FPT}}(S_{1})
    = P(\delta_{r}, S_{r})
    P_{\mathrm{FPT}}(S_{1} \vert \delta_{r}, S_{r}).
\label{eq:backward_proba}
\end{equation}
This equality states that there are two ways to account for trajectories that collapse for the first time at a scale $S_{1}$, \emph{given} that it had the value $\delta_{r}$ at $S_{r}$.
Either one can use the backward probability supplemented by the condition that the crossing at $S_{1}$ was a first crossing [left-hand side of \cref{eq:backward_proba}].
Or one can count the trajectories that from the initial condition $(\delta = 0, S = 0)$, reached the value $\delta_{r}$ at $S_{r}$ without having crossed the barrier [hence $P(\delta_{r}, S_{r})$] and then collapsed at $S_{1}$ from $(\delta_{r}, S_{r})$ [right-hand side of \cref{eq:backward_proba}].

To obtain $\cP(S_{1}, S_{2}; r)$, we schematically start from the first collapsed structure at $(S_1, \deltac(S_1))$, which forms with probability $P_{\mathrm{FPT}}(S_{1})$.
The backward probability allows us to obtain the probability density of the corresponding $\delta_r$ on the vertical slice at $S_r$.
From this point on, we look for the probability to form another PBH at $S_2$ using $P_{\mathrm{FPT}}(S_{2} \vert \delta_{r}, S_{r})$, which can be obtained from \cref{eq:P_FPT_generic}:
\begin{multline}
    \cP(S_{1}, S_{2}; r)
    = \int_{-\infty}^{\deltac(S_{r})} \dd{\delta_{r}}
    P_{\mathrm{FPT}}(S_{1}) \\
    \times P_{\mathrm{bw}}\left[\delta_{r}, S_{r} \vert \deltac(S_{1})\right]
    P_{\mathrm{FPT}}(S_{2} \vert \delta_{r}, S_{r}),
\label{eq:joint_proba_1}
\end{multline}
in which we marginalized over all the possible values of the density contrast $\delta_{r}$ on the vertical slice $S_{r}$.
By construction, the objects formed at $S_1$ and $S_2$ share the same ``history'' before $S_r$, which means that on scales larger than $r$, they have the same realization of the density contrast.

Combining \cref{eq:backward_proba,eq:joint_proba_1} gives
\begin{multline}
    \cP(S_{1}, S_{2}; r) =
    \int_{-\infty}^{\deltac(S_{r})} \dd{\delta_{r}}
    P(\delta_{r}, S_{r}) \\
    \times P_{\mathrm{FPT}}(S_{1} \vert \delta_{r}, S_{r})
    P_{\mathrm{FPT}}(S_{2} \vert \delta_{r}, S_{r}).
\label{eq:joint_probability_final}
\end{multline}
The above expression takes the form of a ``process-convolution kernel''~\cite{higdon2002space} (see also~\cite{genton2001classes,Auclair:2022jod} for a discussion about classes of kernels).
It is symmetric in the exchange of the two PBHs, but cannot be factorized as a product of two functions depending only on $S_{1}$ and $S_{2}$.
This translates to the fact that the two random trajectories are intrinsically correlated: they share a common past.
This is a key difference with the Poissonian approach of PBH clustering, as discussed later in \cref{sec:poisson,sec:comparison}.

\subsection{Marginalized joint probability}

The joint probability given in \cref{eq:joint_probability_final} is relevant when $S_{1}$ and $S_{2}$ are fixed at certain values, that is to say when the masses of the two PBHs under scrutiny are specified, with the constraint that $S_{1}, S_{2} > S_{r}$.
We denote by $P_{2}$ the \textit{marginalized} joint probability which provides the probability of forming two PBHs separated by a distance $r$ without any prior knowledge on their scales
\begin{multline}
    P_{2}
    \equiv \int_{S_r}^{\sigma^{2}} \dd{S_{1}}
    \int_{S_r}^{\sigma^{2}} \dd{S_{2}}
    \int_{-\infty}^{\deltac(S_{r})} \dd{\delta_{r}}
    P_{\mathrm{FPT}}(S_{1} \vert \delta_{r}, S_{r}) \\ \cross
    P_{\mathrm{FPT}}(S_{2} \vert \delta_{r}, S_{r})
    P(\delta_{r}, S_{r}).
\label{eq:P2}
\end{multline}
Note that we integrate $S_n$, $n \in \{1, 2\}$, until $\sigma^{2}$, which serves as a maximum value for $S$ and corresponds to the smallest PBHs.
The lower bound must be $S_{r}$ since, before reaching the scale $S_{r}$, we assumed that there could not have been a collapse, so the two resulting PBHs are necessarily smaller than the distance $r$ from which they are separated from their neighbors.
This last aspect is how the excursion-set approach implements exclusion effects, i.e.\ it is not possible to form two PBHs too close to one another.

\section{Two-point correlation function for fixed threshold}
\label{sec:section3}

\subsection{First-passage-time for fixed threshold}

The formalism exposed in \cref{sec:section2} is fully general and contains all the necessary information to study two-point correlations while incorporating the ``cloud-in-cloud'' mechanism.
They could thus be used in a variety of situations, in particular in complicated cases where the threshold $\deltac(S)$ is scale dependent.

In order to compare our findings with previous literature, we now focus on the case where the formation threshold $\deltac(S)$ is independent of $S$.
This assumption has the other advantage of simplifying greatly the equations and yielding analytical expressions.
In this limit, the Volterra implicit equation \cref{eq:P_FPT_generic} becomes
\begin{equation}
    P_{\mathrm{FPT}} (S \vert \delta_{\mathrm{r}}, S_r) = \frac{\deltac - \delta_{\mathrm{r}}}{\sqrt{2\pi} (S-S_r)^{3/2}} \exp \left[- \frac{(\deltac - \delta_{\mathrm{r}})^{2}}{2(S-S_r)}\right].
\label{eq:P_FPT_cond_fix}
\end{equation}

Similarly, the first-passage-time distribution for trajectories starting from $\delta = 0$ at $S=0$ simply becomes
\begin{equation}
    P_{\mathrm{FPT}}(S) = \frac{\deltac}{\sqrt{2\pi}S^{3/2}} \exp(-\frac{\deltac^{2}}{2S}).
\label{eq:P_FPT_fix}
\end{equation}
Up to a factor of $2$, this is the PBH distribution given by the Press-Schechter formalism.

The distribution $P(\delta, S)$ can be obtained by solving \cref{eq:Fokker-Planck} with an absorbing boundary at $\deltac$
\begin{equation}
    P(\delta_{r}, S_{r}) = \frac{1}{\sqrt{2\pi S_{r}}} \left[\e^{-\frac{\delta_r^{2}}{2S_{r}}} - \e^{- \frac{(2\deltac -\delta_{r})^{2}}{2S_{r}}}\right].
\label{eq:P_fix}
\end{equation}
Thanks to these three expressions, we are able to perform analytical computations for $P_{2}$ and $\cP(S_{1}, S_{2};r)$, and build various quantities to measure spatial correlations.

\begin{figure*}
    \centering
    \subfloat[$\nu = 1$]{\includegraphics[width=0.4\textwidth]{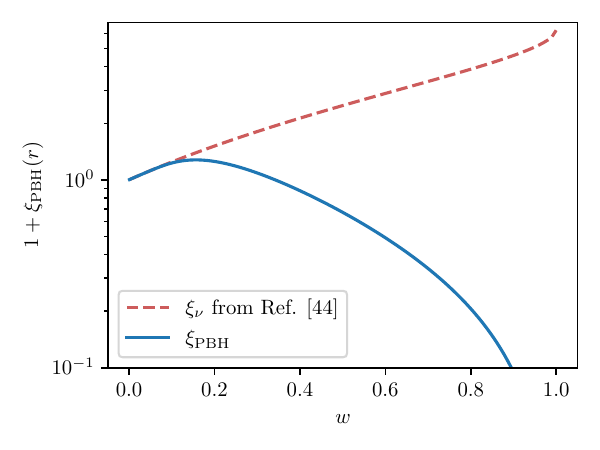}}
    \subfloat[$\nu = 2$]{\includegraphics[width=0.4\textwidth]{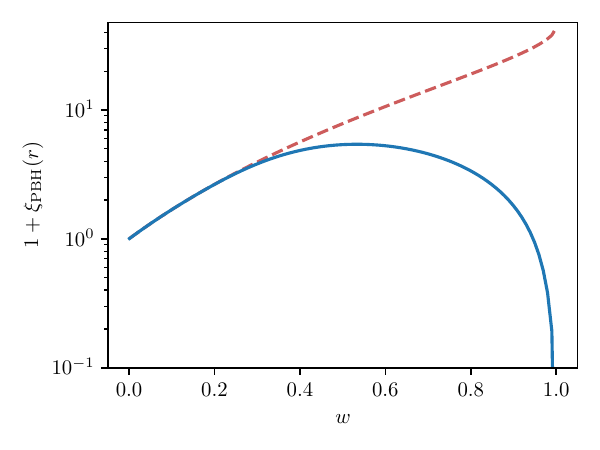}}\\
    \subfloat[$\nu = 5$]{\includegraphics[width=0.4\textwidth]{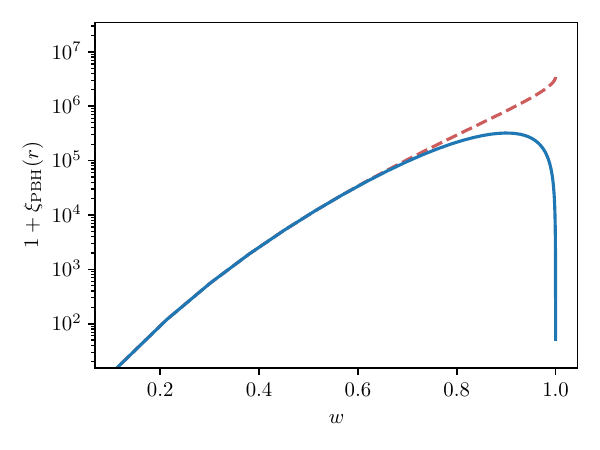}}
    \subfloat[$\nu = 10$]{\includegraphics[width=0.4\textwidth]{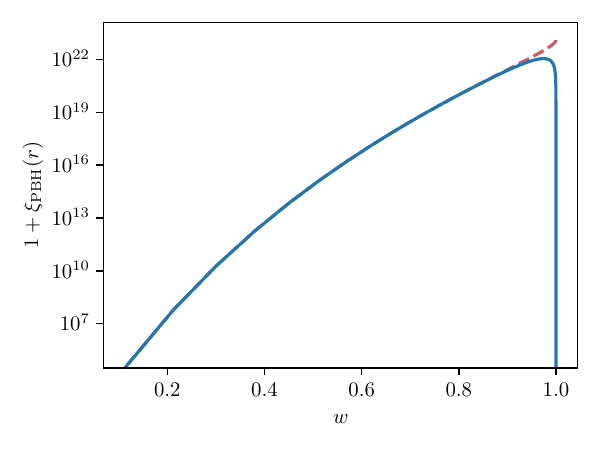}}
    \caption{Two-point correlation function of the PBHs spatial distribution with constant density threshold $\delta_c$, using the Press-Schechter solution of Ref.~\cite{Ali-Haimoud:2018dau} (dashed red) and the Excursion-Set formalism of \cref{sec:section3} (solid blue), see in particular \cref{eq:P2_semianalytic}.}
    \label{fig:Yacine_compar}
\end{figure*}

\subsection{Two-point correlation function}

Following Ref. \cite{Peebles:1980}, the probability $\mathbb{P}(r)$ of finding two objects (irrespective of their nature, e.g.\ galaxies, clusters, although we focus here on PBHs) at separation $r$ is
\begin{equation}
    \mathbb{P}(r) = n^{2} \left[1 + \xi_\mathrm{PBH}(r)\right]
\label{eq:proba_peeble}
\end{equation}
where $n$ is the mean number density of the objects under consideration and is independent of position.
The two-point correlation function $ \xi_\mathrm{PBH}$ measures the excess probability (over random) of finding pairs of objects distant from $r$.
Consistent with homogeneity and isotropy, $ \xi_\mathrm{PBH}$ has been written as a function of the separation alone.

How does \cref{eq:proba_peeble} translate in the excursion-set framework developed in \cref{sec:section2}?
Firstly, we make use of $\sigma^{2}$ introduced in the previous section, which gives the minimum scale of PBH formation.
In the context of PBH, the natural scale $R(S)$ is typically of order of the horizon length at formation: therefore $\sigma^{2}$ corresponds to the value of $S$ when the formation of PBH was first triggered.

The mean number density $n$ is consequently the probability of forming collapsed objects over the range  $\left[0, \sigma^2 \right]$
\begin{equation}
    n = P_{1} \equiv \int_{0}^{\sigma^{2}} P_{\mathrm{FPT}}(s) \dd{s} = \erfc \left(\frac{\nu}{\sqrt{2}}\right) \ ,
\label{eq:prob_P1}
\end{equation}
where we introduced a new variable
\begin{equation}
    \nu \equiv \frac{\deltac}{\sigma} \ ,
    \label{eq:def-nu}
\end{equation}
the formation threshold in units of the standard deviation.

On the other hand, \cref{eq:proba_peeble} is equivalent to the joint probability of \cref{eq:P2}.
Therefore, the excess probability of finding two PBHs within a volume of radius $r$ when ``cloud-in-cloud'' effects are taken into account reads as
\begin{equation}
    1 + \xi_{\mathrm{PBH}}(r) \equiv \frac{P_{2}}{P_{1}^{2}}.
\label{eq:correlation_func_1}
\end{equation}
It is also the mathematical object which is relevant to compare with the correlation function $\xi_{\nu}(r)$ defined in Ref. \cite{Ali-Haimoud:2018dau}, which can be seen as the Press-Schechter equivalent of our $\xi_{\mathrm{PBH}}$.

We emphasize that in $P_{2}$, contrary to $P_{1}$ defined above in \cref{eq:prob_P1}, the condition $S>S_{r}$ is enforced: all the integrals over the variance are bounded by $S_r$ and $\sigma^2$.
This is because the two PBHs under consideration are not independent and thus the volume of radius $r$ in which they are comprised cannot be smaller that the maximum size of the collapse structure formed by each individual PBH.
As we shall discuss later, this is precisely how this excursion-set approach naturally encodes exclusion effects.

For a fixed threshold $\deltac$, the two integrals over $S_{1}$ and $S_{2}$ in $P_{2}$ can be computed analytically.
Using \cref{eq:P_FPT_cond_fix,eq:P_fix} one finds
\begin{equation}
    P_{2} =  \frac{\sqrt{2} \e^{- \frac{\nu^{2}}{2w}}}{\sqrt{\pi w}} \int_{0}^{\infty}  \sinh \left(\frac{\nu}{w}x \right) \erfc^{2} \left[ \frac{x}{\sqrt{2(1-w)}} \right] \e^{- \frac{x^{2}}{2w}} \dd{x}.
\label{eq:P2_semianalytic}
\end{equation}
Using the same notations as in Ref.~\cite{Ali-Haimoud:2018dau}, we introduce the variable $w(r) \equiv S_{r}/\sigma^{2}$ such that $0 \leq w(r) \leq 1$. $w(r)$ approaches zero for infinite separations, and unity for small separations $r$ the size of the smallest PBH.
Furthermore since that $S_r = \langle \delta(\vb{r}) \delta(0) \rangle$ and $\sigma^{2} = \langle\delta^{2}(0)\rangle$, $w(r)$ is nothing else but the two-point correlation function of the underlying density contrast field, normalized by its variance.

From a practical point of view, this form of $P_{2}$ is well suited for numerical evaluation and makes $\xi_{\mathrm{PBH}}(r)$ easy to study.
We display our result for $\xi_\mathrm{PBH}$ in \cref{fig:Yacine_compar}, along with the Press-Schechter-inspired model of Ref.~\cite{Ali-Haimoud:2018dau}.
The trend exhibited by the correlation function of Ref.~\cite{Ali-Haimoud:2018dau}, that we shall denote by $\xi_{\nu}$ hereafter, is recovered at large separations (i.e.\ $w \to 0$) whereas significant deviations are obtained at small scales.

The importance of these deviations as well as the range of $w$ for which they occur increases as $\nu$ becomes small.
This can be understood in regard to the physical meaning of $\nu$.
As explicit in \cref{eq:def-nu}, $\nu$ is the ratio between $\deltac$ and $\sigma$, and as illustrated in \cref{fig:random_bundle}, it is a measure of the barrier's ``height'', in units of the typical dispersion of the random trajectories.
High values of $\nu$ represent a formation threshold which is far from the region where trajectories are typically spread, and thus barrier's crossings can be seen as relatively ``rare'' events. Conversely, low values of $\nu$ bring $\deltac$ closer to the region where fluctuations are the most concentrated: multiple crossings are much more frequent and this is thus the situations where the ``cloud-in-cloud'' mechanism is expected to play its most significant role.

Bearing this in mind, we outline that $\xi_{\mathrm{PBH}}$ departs from $\xi_{\nu}$ in three main ways. First, $\xi_{\mathrm{PBH}}$ is always below $\xi_{\nu}$ for all values of $w$, meaning that the result from Ref. \cite{Ali-Haimoud:2018dau} systematically overestimates the PBH clustering. This overestimation is particularly significant for low values of $\nu$. As discussed just above, it translates the effect of the ``cloud-in-cloud'' mechanism that was missed in Ref.~\cite{Ali-Haimoud:2018dau} and in the previous literature, since multiple crossing of the threshold is more probable for low values of $\nu$.

Second, $\xi_{\mathrm{PBH}}$ systematically exhibits a maximum, which is reached at higher values of $w$ as $\nu$ increases, thus defining a typical correlation length $w_{\xi_{\mathrm{PBH}}}$. Due to the form of the derivative of $P_{2}$ with respect to $w$ no analytical expression of this correlation length can be obtained. However, we see that it is highly sensitive to $\nu$ (see \cref{fig:Yacine_compar}).

Third, as $w \to 1$, the excess probability $1 + \xi_{\mathrm{PBH}}$ drops to zero contrary to $\xi_{\nu}$ which reaches the bound $2/P_{1}$ for $w=1$. More generally, our excursion-set formalism provides a drastic difference of behavior at small-scale in comparison to previous works, and this question is the subject of an in-depth discussion which we now address.

\section{Small-scale behavior: exclusion effects and shot noise}
\label{sec:small}

On sufficiently large scales, the behavior of the PBHs is expected to follow the one of standard adiabatic perturbations. However, on small enough scales, their discrete nature becomes important \cite{Carr:2018rid, Inman2019}, which is the reason why in the literature, a specific emphasis was put on the small-scale behavior of PBHs.

\subsection{Volume-exclusion effects}

In all previous approaches pertaining to their clustering, PBHs were treated as pointlike (see e.g.\ \cite{Chisholm:2005vm, Ali-Haimoud:2018dau}). To account for their finite size, some authors argued the necessity to add, in a second time, spatial exclusion effects \cite{Mo:1995cs, Baldauf:2015fbu, Desjacques:2016bnm}.
This spatial exclusion is usually enforced by introducing by hand a lower bound $r_{\mathrm{exc}}$ on $r$ which is, in the context of PBH formation, of the order of a few horizon lengths \cite{Desjacques:2018wuu}. Then the condition $\xi_{\mathrm{PBH}}(r) \approx -1$ for $r \lesssim r_{\mathrm{exc}}$ is artificially imposed.

We emphasize that in our approach, this procedure is spurious since that, as \cref{fig:Yacine_compar} clearly shows, exclusion effects are already contained in our expression of $\xi_{\mathrm{PBH}}$. Indeed, by design, we have $S_r < S$, which eventually means we only look for separations such as $r > R(S)$ being the size of the Hubble patch of the subsequently formed PBH. Mathematically, the limit $w \to 1$ is analytical in \cref{eq:P2_semianalytic} for a fixed threshold, and also apparent for any threshold $\deltac(S)$ in \cref{eq:P2} when $S_r \to \sigma^2$.
Ultimately this gives
\begin{equation}
    \lim_{w \to 1^-} \xi_{\mathrm{PBH}}(w) = -1 \ .
\end{equation}
Thus PBHs are anticorrelated at short distances. As a consequence, our formalism enforces that PBHs cannot form arbitrarily close to each other, thus effectively including short-range exclusion effects. As pointed out in Refs. \cite{Desjacques:2018wuu, Baldauf:2015fbu}, our expression of $\xi_{\mathrm{PBH}}$ also stands as a counterexample to the claim of Ref. \cite{Ali-Haimoud:2018dau} that the joint probability should go to unity when $r \to 0$, since in our case, $P_{2} \to 0$ in this limit.

Let us also notice that exclusion effects spread for $w \lesssim 1$; this phenomenon is impossible to describe when imposing a hard cutoff such as in~\cite{Desjacques:2018wuu}.
In \cref{fig:Yacine_compar}, we show that PBHs are still anticorrelated for $w \lesssim 1$, and the range of values of $w$ over which this anticorrelation spreads gets wider as $\nu$ decreases. For instance, for $\nu = 1$, PBHs are nearly always anticorrelated.
This proves that the ``cloud-in-cloud'' mechanism can significantly enhance spatial exclusions and thus anticorrelation of PBHs, sometimes well beyond the short-range region.

\subsection{Poisson shot noise}

With regard to their macroscopic mass, it was originally convenient to think of the PBH distribution as a distribution of pointlike objects, and by analogy, employing the tools that were already developed for the study of galaxies and clusters of galaxies. A consequence of such a discrete treatment is the appearance, at $\vb{r} = 0$, of a Poisson shot noise modeling the ``self-pairs'' contribution \cite{Peebles:1980, Baldauf:2013hka}. This idea was first suggested in Ref. \cite{Meszaros1974, Meszaros:1975ef} where the existence, in any given volume, of a statistical fluctuation in the number of initially formed PBHs was pointed out. This fluctuation is coming only from the discrete nature of PBHs, irrespectively of their formation process. It was later shown in~\cite{Carr1977} that for random uncorrelated number fluctuations, this statistics was necessarily of Poissonian nature, i.e.\ the fluctuation
$\delta N /N \propto 1/\sqrt{N}$ if $N$ is the number of PBHs.

\subsection{Poissonian clustering}
\label{sec:poisson}

Subsequent developments approximate such a random (Poisson) \textit{point} process by a \textit{continuous} approach \cite{Freese1983, Carr1983, Afshordi:2003zb}.
As explained in~\cite{Limber1957,Peebles:1980}, one postulates the existence of a continuous probability density field $\rho_{\mathrm{PBH}}$, whose mean corresponds to the number density $n$ of PBHs and with a nonzero two-point correlator.
For a volume element $\delta V$ centered on a position $\vb{r}$, a PBH is found with probability
\begin{equation}
    \rho_{\mathrm{PBH}}(\vb{r}) \delta V.
\end{equation}
In this approach, the joint probability to find two objects in the volumes $\delta V_{1}$ and $\delta V_{2}$ is simply
\begin{equation}
    \rho_{\mathrm{PBH}}(\vb{r}_{1}) \rho_{\mathrm{PBH}}(\vb{r}_{2})\delta V_{1} \delta V_{2}.
    \label{eq:poisson}
\end{equation}
In the theory of kernels, this joint probability takes the form of a ``separable kernel''~\cite{genton2001classes}, i.e.\ it is the product of the same function for objects $1$ and $2$.
Although the probability of finding a PBH in $\delta V_{1}$ is by construction independent to the one of finding a PBH in $\delta V_{2}$, the two-point correlator of $\rho_{\mathrm{PBH}}$ is nonvanishing.
Indeed, the field $\rho_{\mathrm{PBH}}$ that underlies the distribution of PBHs can depend on space and thus induces spatial correlations
\begin{equation}
    \xi(\vb{r}) \equiv \frac{\ev{[\rho_{\mathrm{PBH}}(\vb{r}_{1}) - n] [\rho_{\mathrm{PBH}}(\vb{r}_{2}) - n]}}{n^2} \neq 0.
\end{equation}
Consequently, in such a framework, sometimes referred as Poisson model \cite{Peebles:1980}, the correlation function $\xi(\vb{r})$ can be introduced and derived from $\rho_{\mathrm{PBH}}$.

As such, $\xi$ is a descriptive statistic, since $\rho_{\mathrm{PBH}}$ is formally unrelated to the underlying physical radiation density contrast $\delta$ actually forming the PBHs (although this is the aim of bias theory, see e.g.\ \cite{Desjacques:2016bnm} for a review).
It is thus defined as a model -- presenting convenient properties -- but, for instance, as noted in~\cite{Peebles:1980}, it cannot describe a process for which $\xi = -1$ under a certain distance $r$.
This is the reason why in this approach, introducing a cutoff to implement anticorrelation at short distance is necessary.

This continuous approach is expected to reproduce the Poisson fluctuations at small scales.
Indeed, $\rho_{\mathrm{PBH}}$ also incorporates a (Poisson) shot noise in the number density of PBHs, arising from their discrete nature.
On top of this shot noise, the two-point function of $\rho_{\mathrm{PBH}}$ describes the amount of additional clustering that can be present~\cite{Desjacques:2018wuu}.

\subsection{Comparison with our framework}
\label{sec:comparison}

This brief review highlights that our formalism differs from the above approach in two aspects.

First and foremost, we do not treat PBHs as pointlike objects.
They have a finite size, and their spatial extension does not allow the two PBHs to be located arbitrarily close.
As a consequence, $\xi_{\mathrm{PBH}}$ vanishes at small separations and is not concerned by the ``zero-lag'' limit and its associated Poisson noise invoked in previous works.
Let us however note that ``zero-lag'' effects may still appear when one studies individual pairs of PBHs with a given mass, as we discuss later in \cref{sec:section4}.

Second, when examining two PBHs separated by a certain distance, we do not assume that their formation is completely uncorrelated.
On the contrary, since their random trajectories share a common past (from $S=0$ to the scale $S_{r}$), it induces correlations which cannot be taken into account with the Poisson model of clustering.

Indeed, as briefly discussed in \cref{sec:joint}, this fundamental difference between Poisson clustering and our result can be interpreted in the theory of kernels (see~\cite{genton2001classes,Auclair:2022jod} for reviews).
In short, kernels are objects that encode the properties of two-point correlators.
As such, kernels cannot be any function of two variables and need to satisfy a number of mathematical properties, i.e.\ symmetry, positive definiteness and Mercer's condition~\cite{mercer1909xvi}.
In the context of this paper, the kernel $K_{S_r}(S_1, S_2)$ is the probability of finding two PBHs of masses $(S_1, S_2)$ separated by $S_r$.
The Poissonian models giving \cref{eq:poisson} assume that this kernel is ``separable'': there exists a function $f(S)$ such that
\begin{equation}
    K_{S_r}(S_1, S_2) = f(S_1) f(S_2).
\end{equation}

To the contrary, our prediction using the excursion-set formalism of \cref{eq:joint_proba_1} pertains to a wider class of kernels dubbed as ``process-convolution kernels''~\cite{higdon2002space}.
That is, there exists a function $g(S, \delta_r)$ such that
\begin{equation}
    K_{S_r}(S_1, S_2) = \int g(S_1, \delta_r) g(S_2, \delta_r) \dd{\delta_r}.
\end{equation}
Process-convolution kernels provide a richer phenomenology than separable kernels and have proven useful in various fields of physics ranging from hydrology and atmospheric science~\cite{higdon2002space}, to machine learning~\cite{paciorek2006spatial} and freely decaying turbulence in the early Universe~\cite{Auclair:2022jod}.

\section{Pairwise correlation functions for fixed threshold}
\label{sec:section4}

\begin{figure*}
    \centering
    \subfloat[$\lambda = 1$]{\includegraphics[width=0.45\textwidth]{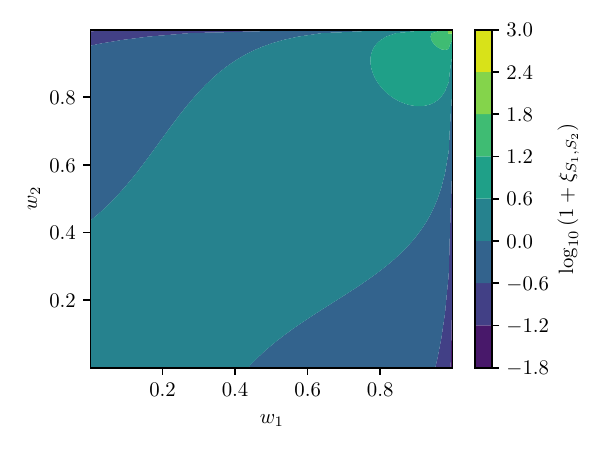}}
    \subfloat[$\lambda = 2$]{\includegraphics[width=0.45\textwidth]{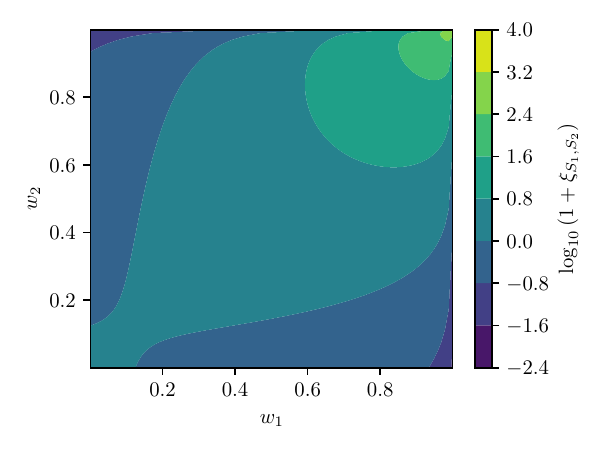}}\\
    \subfloat[$\lambda = 5$]{\includegraphics[width=0.45\textwidth]{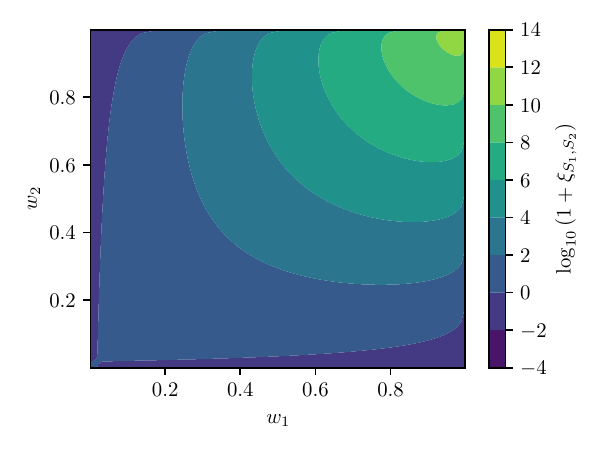}}
    \subfloat[$\lambda = 10$]{\includegraphics[width=0.45\textwidth]{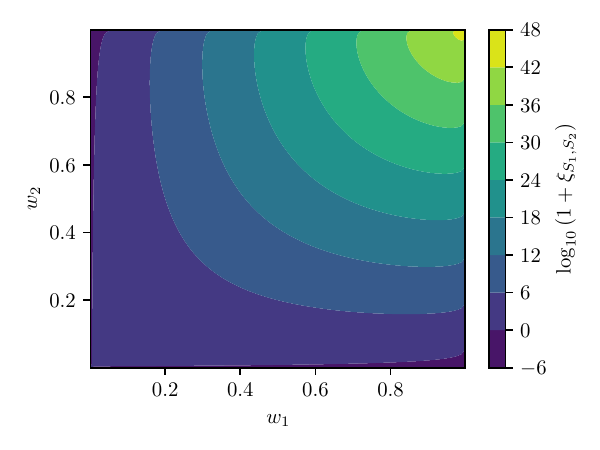}}
    \caption{Excess probability to find pairs of PBHs with masses $S_1, S_2$ at a fixed distance $r$.
    $w_n = S_r / S_n$ is a measure of the PBH mass, the limit $w_n \to 0$ corresponds to a very small PBH, whereas $w_n \to 1$ corresponds to the maximum allowed PBH mass in the volume contained within the two PBHs.}
\label{fig:Contour_plots}
\end{figure*}

The correlation function $\xi_{\mathrm{PBH}}(r)$ examined in the previous section provides global information on the spatial clustering, as it is integrated over a broad mass range.
From an observational standpoint, it is indeed what is typically measured since the mass of the observed objects (whether galaxies, stars, black holes, etc.) is \textit{a priori} unknown.
However, our framework is well suited to go beyond and to study correlations directly at the level of pairs of PBHs with given masses.
Although we still remain model independent, i.e.\ we do not specify the power spectrum, interesting conclusions can be drawn.

In this so-called ``pairwise'' approach, we make use of the joint probability $\cP(S_1, S_2;r)$ of \cref{eq:P2}.
In similar fashion as in \cref{eq:proba_peeble}, we define the cross-correlation function $\xi_{S_1,S_2}(r)$ as the excess probability (over random) of finding a PBH of mass $M_{1}$ and a PBH of mass $M_{2}$ distant from $r$
\begin{multline}
    1 + \xi_{S_1,S_2}(r)
    = \frac{\cP(S_{1}, S_{2} ; r)}{P_{\mathrm{FPT}}(S_{1}) P_{\mathrm{FPT}}(S_{2})} \\
    = \int_{-\infty}^{\deltac} \dd{\delta_{r}} P(\delta_{r}, S_{r})
    \frac{P_{\mathrm{FPT}}(S_{1} \vert \delta_{r}, S_{r})  P_{\mathrm{FPT}}(S_{2} \vert \delta_{r}, S_{r})}{P_{\mathrm{FPT}}(S_{1})P_{\mathrm{FPT}}(S_{2})} .
\end{multline}
As expected, let us notice that the above quantity is symmetric in the exchange of $S_{1}$ and $S_{2}$.
Taking the limit $S_{1} \to S_{2}$ provides $\xi_{S,S}(r)$ which counts, given a population of PBHs of mass $M$, the excess probability of finding another member with same mass $M$ at distance $r$:
\begin{equation}
    1+\xi_{S,S}(r) = \int_{-\infty}^{\deltac} \dd{\delta_r} \left[ \frac{P_{\mathrm{FPT}}(S \vert \delta_{r}, S_{r})}{P_{\mathrm{FPT}}(S)} \right]^{2} P(\delta_{r}, S_{r}).
\label{eq:xi_SS}
\end{equation}
This is the first-crossing equivalent of the correlation function $\xi_{MM}(r)$ in Ref.~\cite{Hopkins:2012bi, Hopkins:2012fm} in the context of stellar spatial clustering (with last-passage-times).

Making use of \cref{eq:P_FPT_cond_fix,eq:P_FPT_fix,eq:P_fix} as well as the following integral from Ref.~\cite{gradshteyn2007}
\begin{multline}
    \int_{0}^{\infty} \dd{x} x^{2} \e^{-\beta x^{2}} \sinh (\gamma x) \\
     = \frac{\sqrt{\pi} (2\beta + \gamma^{2})}{8 \beta^{2} \sqrt{\beta}} \e^{\frac{\gamma^{2}}{4\beta}} \erf \left(\frac{\gamma}{2 \sqrt{\beta}} \right)
     + \frac{\gamma}{4\beta^{2}},
\end{multline}
the cross-correlation function can be computed in a fully analytical way
\begin{widetext}
\begin{equation}
\begin{split}
    1 + \xi_{S_1,S_2}(r) = &\frac{\e^{\lambda^{2}(w_{1} + w_{2}-1)}}{\sqrt{\pi}\lambda} \frac{\sqrt{(1-w_{1})(1-w_{2})}}{(1-w_{1}w_{2})^{2}} \Bigg\{ 1+ \sqrt{\pi} \lambda  \sqrt{\frac{(1-w_{1})(1-w_{2})}{1 - w_{1}w_{2}}} \\
    &\times \left[\frac{1}{2\lambda^{2}} \frac{1 - w_{1}w_{2}}{(1-w_{1})(1-w_{2})} +1 \right] \e^{\lambda^{2} \frac{(1-w_{1})(1-w_{2})}{1 - w_{1}w_{2}}} \erf \left[\lambda \sqrt{\frac{(1-w_{1})(1-w_{2})}{1 - w_{1}w_{2}}} \right]    \Bigg\}.
\end{split}
\label{eq:cross_correlation_application}
\end{equation}
\end{widetext}
where we introduced the parameters $w_{n}(r) = S_{r}/S_{n} \in ]0, 1[$, $n \in \{1, 2\}$, and
\begin{equation}
    \lambda \equiv \frac{\deltac}{\sqrt{2S_{r}}} \ .
\end{equation}
Note that the explicit expression of \cref{eq:xi_SS} can be straightforwardly obtained by taking the limit $S_{1}, S_{2} \to S$, that is $w_{1}, w_{2} \to w$ in \cref{eq:cross_correlation_application}.

Unlike \cref{sec:section3}, we are now specifying the masses of the two PBHs, therefore our result does not depend on the cutoff scale $\sigma^2$ anymore.
This is why we introduce a new quantity $\lambda$, whose definition resembles $\nu$ but where the reference scale $\sigma^2$ is replaced by $S_r$.
As a consequence of $\lambda$ depending on $S_r$, \cref{eq:cross_correlation_application}  should be interpreted as information on what kind of mass hierarchy is favored by spatial clustering, at a fixed distance $r$, i.e.\ at fixed $\lambda$.

To interpret $\xi_{S_1,S_2}(r)$, we fix the distance between the pair of PBHs (i.e.\ $S_{r}$) as well as the height of the barrier (i.e.\ $\deltac$) so that the parameter $\lambda$ is entirely specified.
We let $(w_{1}, w_{2})$ vary, that is to say the mass of the two PBHs.
The contour plots for different values of $\lambda$ resulting from this approach are displayed in \cref{fig:Contour_plots}.
First, they show that pairs of PBHs with high masses are always more clustered than their lighter counterparts.
Second, we can observe that the contour plots present relatively broad aspects, instead of being concentrated in the diagonal for which $w_{1} \sim w_{2}$.
It means that the clustering between pairs of PBHs of relatively similar masses is not favored in comparison to pairs of PBHs presenting a high hierarchy of masses i.e.\ for which $w_{1} \ll w_{2}$ or conversely.
This result is most significant for high values of $\lambda$ but is still present for $\lambda = 1,2$.

Finally, let us notice that $\xi_{S_1,S_2}(r)$ does experience some sort of ``zero-lag'', as can be seen in the upper right corners $(w_1, w_2) \to (1^-, 1^-)$ of \cref{fig:Contour_plots}.
The interpretation for this peak as follows.
Taking a region of space of size $S_r$, we impose the formation of one PBH $S_1$ centered around $\vb{x}_1$ which takes almost all the available space in the volume, i.e.\ $w_1 \to 1^-$.
Most choices of $\vb{x}_2$ in the volume of size $S_r$ are most likely included in the same PBH with mass $S_2 \approx S_1$.
This explains why the excess probability $\xi_{S_1,S_2}(r)$ converges to a Dirac distribution centered around $S_2 \to S_1$ on the vertical slice $w_1 \to 1^-$.
The excursion-set approach is valid provided a certain separation of scales between the volume $S_r$ and the size of the collapsed regions in order to perform ensemble averages.
This small scale behavior for given pairs of PBHs is expected and allows us to show the limit of applicability of our work, that is \cref{fig:Contour_plots} suffers from zero-lag and is invalid in the vicinity of $w_n \sim 1$.
Nonetheless, our conclusions on the mass ratio of binaries are still perfectly applicable.

\section{Conclusion}
\label{sec:section5}

In this work, we used the excursion-set formalism to investigate the initial small-scale spatial clustering of PBHs.
The aim was to go beyond the Press-Schechter-inspired approaches proposed in earlier works.
By employing first-passage-time methods, accounting for the ``cloud-in-cloud'' mechanism, and by reliably distinguishing the masses of the PBHs, we were able to compute a two-point correlation function $\xi_{\mathrm{PBH}}$ featuring novel interesting properties.
In particular, we find $\xi_{\mathrm{PBH}} \to -1$ at small separations, i.e.\ PBHs are anticorrelated at short distance.
Since we account for the finite size of the PBHs, it effectively reproduces volume-exclusion effect, which stands as an improvement from the simplified pointlike approach.

While the quantity $\xi_{\mathrm{PBH}}$ examines correlations of a population of PBHs irrespective of their masses, we also derived a novel ``pairwise'' approach at the level of pairs of PBHs, when their masses are known and specified.
This new approach highlights that the clustering between pairs of PBHs of relatively similar masses is not particularly favored in comparison to pairs of PBHs presenting a high hierarchy of masses.

For these two objects, in the context of PBHs formed out of the Hubble reentry of large density fluctuations, we obtained \textit{exact} analytical expressions when considering the case of a scale-invariant threshold as the criterion of formation.
However, our formalism is generic: the joint probabilities we introduced, and which contain all the necessary information to study two-point correlations, can also be used in full generality, without any constraint on the density threshold.

\bigskip

Throughout this work, we decided to remain as model independent as possible pertaining to the PBH process of formation. Thus, the new tools introduced herein could be used in various situations. In particular, our new approach is well suited to study PBH formation models with broad power spectra, in which case $S(R)$ becomes a smooth function of the scale $R$, and when PBH formation happens frequently, that is $\nu, \lambda = \order{1}$.
To put things into perspective, the fractional volume contained into PBHs for a scale-invariant threshold is given by~\cite{Auclair:2020csm}
\begin{equation}
    f_{\mathrm{PBH}, V} = \erfc\left(\frac{\nu}{\sqrt{2}}\right).
\end{equation}
That is in \cref{fig:Yacine_compar}, $f_{\mathrm{PBH}, V}$ ranges from $0.31$ for $\nu = 1$, to $0.045$ for $\nu = 2$ and to $5.7\times 10^{-7}$ already for $\nu = 5$.
Therefore, we expect this exclusion effect to have the biggest impact when PBH formation is abundant, but also to leave a sizeable signature when PBH formation is infrequent.

The pairwise approach could also have exciting uses.
First, it enables to study correlations in the neighborhood of a given PBH, and analyze whether a massive PBH has an excess probability of being surrounded by smaller PBHs, or the converse.
Second, it could be used to better assess PBHs merging rates.
The number of mergers is indeed quite sensitive to the initial clustering of PBHs (and its subsequent nonlinear evolution) as well as the preferred mass ratios~\cite{DeLuca:2020jug}.

\bigskip

The fact that we remained model independent has two main consequences.

First, no quantitative estimations of amounts of clustering were performed in this work.
Indeed, applications would require a detailed relativistic criterion for the formation of PBHs, a model for the power spectrum of the density fluctuations and a thorough discussion on certain choices of gauges for the cosmological perturbations.
For instance, an interesting scenario that could be worth investigating -- and where it was already proven that it was significantly affected by ``cloud-in-cloud'' mechanism \cite{Auclair:2020csm} -- would be PBHs arising from preheating instabilities.

The second consequence is that we do not have access to real-space distances, since $S(R)$ is left unspecified.
This leads to an imperfect ``exclusion-effect'' on small scales.
Ideally, one would impose that the sum of the two Hubble patch radii $r_1, r_2$ should be less than the distance $r$ between the PBHs, i.e.\ $r_1 + r_2 < r$.
However, we are only able to impose $r_1 < r$ \textit{and} $r_2 < r$ if we want to remain model independent.
As discussed extensively, this approximation is sufficient to induce an effective exclusion effect on $\xi_\mathrm{PBH}$, but we showed in \cref{sec:section4} that the limit of this approach is reached at ``zero-lag'', in the vicinity of $S \approx S_r$, that is the upper right corners of \cref{fig:Contour_plots}.

Eventually, the most salient prospect of improvement of our formalism would be to implement, on top of the ``cloud-in-cloud'' mechanism, non-Gaussian initial conditions. The non-Markovianity of the random walks induced by non-Gaussianity could have huge impacts on the two-point correlation function.
The correlated steps' problem for the computation of halo mass functions in excursion-set theory was already investigated in Refs. \cite{Maggiore:2009rv, Maggiore:2009rw, Maggiore:2009rx}, extended to two-point statistics in the context of the two-barrier setup in Ref. \cite{DeSimone:2011dn} and later generalized -- although differently -- in Refs. \cite{Musso:2011ck, Musso:2012ch, Musso:2012qk, Musso:2013pha, Musso:2013pja}.
However, we believe that new ways of accounting for non-Gaussian effects combined with our present framework could be developed thanks to the stochastic inflation formalism \cite{Starobinsky:1986fx, Goncharov:1987ir, Starobinsky:1994bd, Vennin:2015hra, Grain:2020wro}.

\begin{acknowledgments}
We kindly thank Christophe Ringeval, Vincent Vennin, Chiara Animali and Julien Lavalle for fruitful discussions and for their precious feedback on the first version of the manuscript.
We also thank Cristiano Germani and Albert Escriv\`a for their comments on a later version of the manuscript.
The work of P.A. is supported by the Wallonia-Brussels Federation Grant ARC NO. 19/24-103.
B.B. is publishing in the quality of ASPIRANT Research Fellow of the ``Fonds de la Recherche Scientifique - FNRS''.
\end{acknowledgments}

\bibstyle{aps}
\bibliography{biblio}

\end{document}